\documentclass[aps,preprint,epsfig,rotate]{revtex4}
\usepackage{graphicx}
\usepackage{bm}
\usepackage{epsfig}

\input epsf
\begin{document}
\title{Bound state spectra and properties of the five- and six-body muonic ions and quasi-molecules.}

\author{Alexei M. Frolov}
\email[E--mail address: ]{afrolov@uwo.ca}

\affiliation{ITAMP, Harvard-Smithonian Center for Astrophysics, \\
         MS 14, 60 Garden Street, Cambridge MA 02138-1516, USA}  

\affiliation{Department of Applied Mathematics \\
       University of Western Ontario, London, Ontario N6H 5B7, Canada} 

\date{\today}

\begin{abstract}
The ground bound states in the five-body muonic ions $a b \mu e_2$ (or $(a b \mu e_2)^{-}$), where $(a, b) = (p, d, t)$, are considered for the first time. As follows 
from accurate numerical computations of these five-body ions they are similar to the negatively charged hydrogen ion H$^{-}$ with a three-particle central quasi-nucleus 
$a b \mu$ (or $(a b \mu)^{+}$). These five-body ions play some role in the muon-catalyzed fusion of nuclear reaction in liquid deuterium and/or deuterium-tritium 
mixtures. We also discuss the bound states in the six-body $a b c \mu e_2$ quasi-molecules, where $(a, b, c) = (p, d, t)$. These six-body, two-electron 
quasi-molecules are of great interest for muon-catalyzed nuclear fusion, since in liquid deuterium  and/or deuterium-tritium mixture more than 99 \% of all 
muon-catalyzed reactions of the $(dd)-$ and $(dt)-$fusion occur in such six-body quasi-molecules: $d d t \mu e_2, d d d \mu e_2$ and $d t t \mu e_2$.  

\noindent 
PACS number(s): 36.10.Dr


\end{abstract}

\maketitle
\newpage

\section {Introduction}

The goal of this study is to analyze the bound state spectra and internal structure of a number of five- and six-body species which contain one negatively charged muon 
$\mu^{-}$ and two bound electrons $e^{-}$. These compounds are formed when fast negatively charged muons $\mu^{-}$ slow down in liquid deuterium (hydrogen) and during 
the following chain of reactions of muon-catalyzed fusion. In various papers published since the end of 1950's (see, e.g., \cite{Sachar} - \cite{Advc}) it was assumed 
that the reactions of the nuclear $(d,d)-$ and $(d,t)-$fusion 
\begin{eqnarray}
  & & d + d = {}^{3}{\rm He} + n + 3.270 MeV \; \; \; ; \; \; \;  d + d = t + p + 4.033 MeV \label{eq1} \\
  & & d + t = {}^{4}{\rm He} + n + 17.590 MeV \; \; \; \label{eq2}
\end{eqnarray}
proceed in the three-particle muonic molecular ions $dd\mu$ and $dt\mu$, respectively (see, e.g., \cite{MS}). It is clear that the actual fusion rates substantially 
depend upon the quantum numbers of actual bound state(s) in three-body muonic molecular ions $pp\mu, pd\mu, pt\mu, dd\mu, dt\mu$ and $tt\mu$. The total energies of all 
22 bound states in six muonic molecular ions $pp\mu, pd\mu, pt\mu, dd\mu, dt\mu$ and $tt\mu$ can be found in \cite{Fro2011}. In reality, it is difficult to imagine that 
such compact ions can exist as bare positively charged quasi-particles at very low temperatures which can be found in liquid hydrogen, or in liquid deuterium. In general, 
all three-particle muonic molecular ions will attract a number of electrons from surronding molecules. In addition to this, the positively charged $dd\mu$ and $dt\mu$ 
ions and neutral hydrogen-like $d d \mu e$ and $d t \mu e$ atoms will interact with the atoms of deuterium, tritium and hydrogen molecules from liquid deuterium 
and/or deuterium-tritium mixture. Such interactions will lead, in principle, to the formation of a number of four-, five- and six-particle atomic and molecular muonic 
systems, i.e. systems which include one negatively charged muon $\mu^{-}$, two bound electrons $e^{-}$ and two (or three) nuclei of hydrogen isotopes. 

In this study we consider the five-particle atomic ions, such as $p d \mu e_2, d d \mu e_2, d t \mu e_2$, etc, and six-body neutral quasi-molecules, such as $p d t \mu e_2$ 
and $d d t \mu e_2$. Based on our current knowledge of internal structures of various few-body atoms and ions we can predict that all five-body, two-electron 
$(a b \mu e_2)^{-}$ ions are similar to the negatively charged hydrogen ion H$^{-}$, which has only one stable (ground) electronic $1^1S$-state. Here and everywhere below
the letters $a, b$ and $c$ designate the nuclei of hydrogen isotopes. In contrast with this, the internal structure of the six-body quasi-molecules $p d t \mu e_2, 
d d t \mu e_2$ is similar to the structure of the neutral two-electron H$_{2}$ molecule. As was found in actual experiments more than 99 \% of all $(dd)-$ and $(dt)-$fusion 
reactions in liquid deuterium and/or deuterium-tritium mixture occur in such six-body quasi-molecules: $p d t \mu e_2, d d d \mu e_2$ and $d d t \mu e_2$ (see, e.g., 
\cite{MS} and references therein). In this study we investigate the basic properties and internal structure of these six-body quasi-molecules $a b c \mu e_2$. It should be 
mentioned here that none of these five- and six-body muonic ions and quasi-molecules have been considered earlier. 

Our goal below is to perform variational computations of the ground $S(L = 0)-$states in the five-body ions $a b \mu e_2 = (a b \mu e_2)^{-}$, where $(a, b) = (p, d, t)$, 
and in the six-body $a b c \mu e_2$ quasi-molecules, where $(a, b, c) = (p, d, t)$. Briefly, we consider the bound ground $S(L = 0)-$states in a number of five- and six-body 
two-electron systems each of which contains one negatively charged muon $\mu^{-}$, two electrons $e^{-}$ and two (or three) bare nuclei of hydrogen isotopes, i.e. the nuclei 
of protium $p$, deuterium $d$ and/or tritium $t$. In our computations we determine the total energies and basic geometrical and physical properties of these ions and 
quasi-molecules. All calculations in this work have been performed with the use the following particle masses:
\begin{eqnarray}
  & &m_{\mu} = 206.768262 m_e \; \; \; , \; \; \;  M_p = 1836.152701 m_e \label{eq01} \\
  & &M_d = 3670.483014 m_e \; \; \; , \; \; \; M_t = 5496.92158  m_e \nonumber
\end{eqnarray}
where $m_e$ is the electron's mass. The same set of particle masses has been used in our calculations of the three-body muonic molecular ions \cite{Fro2011}. In this study 
we apply the atomic units in which $\hbar = 1, e = 1$ and $m_e = 1$. Our goal below is to determine the accurate solutions of the non-relativistic Schr\"{o}dinger equation 
for the discrete spectrum $H \Psi = E \Psi$ (where $E < 0$). The non-relativistic Hamiltonians of the five- and six-body systems considered in this study are
\begin{eqnarray}
  H_5 = -\frac{\hbar^{2}}{2 m_e} \Bigl[\nabla^2_{1} + \nabla^2_{2} + \frac{m_e}{m_{\mu}} \nabla^2_{\mu} + \frac{m_e}{M_a} \nabla^2_{a} + \frac{m_e}{M_b} \nabla^2_{b} \Bigr] + 
 \sum^{5}_{i=2} \sum^{4}_{j=1 (j<i)} \frac{q_i q_j e^{2}}{r_{ij}} \; \; \; , \; \; \label{Hamilt5}
\end{eqnarray}
and 
\begin{eqnarray}
  H_6 = -\frac{\hbar^{2}}{2 m_e} \Bigl[\nabla^2_{1} + \nabla^2_{2} + \frac{m_e}{m_{\mu}} \nabla^2_{\mu}  + \frac{m_e}{M_a} \nabla^2_{a} + \frac{m_e}{M_b} \nabla^2_{b} + 
 \frac{m_e}{M_c} \nabla^2_{c} \Bigr] + \sum^{6}_{i=2} \sum^{5}_{j=1 (j<i)} \frac{q_i q_j e^{2}}{r_{ij}} \; \; \; , \; \; \label{Hamilt6}
\end{eqnarray}
respectively. Here and everywhere below in this study we assume that the indexes 1 and 2 designate the two bound electrons, the index 3 stands for the negatively charged muon 
$\mu^{-}$, while indexes 4, 5 and 6 denote the two (or three) heavy hydrogen nuclei. In atomic units the explicit forms of these two Hamiltonains, Eqs.(\ref{Hamilt5}) - 
(\ref{Hamilt6}), are drastically simplified, since in this case in Eqs.(\ref{Hamilt5}) - (\ref{Hamilt6}) we have $e^2 = 1, \hbar^2 = 1$ and $m_e = 1$.

This paper has the following structure. Variational computations of the five-body ions $a b \mu e_2$, where $(a, b) = (p, d, t)$, are considered in the next Section. Here we 
also compare some of the bound state properties of these ions with the known expectation values for the negatively charged hydrogen ion(s) H$^{-}$ \cite{Fro2015}. Numerical 
calculations of the total energies and a few other bound state properties of the six-body quasi-molecules $a b c \mu e_2$, where $(a, b, c) = (p, d, t)$, are performed in 
Section III. Comparison of these properties with the known properties of the H$_2$ molecule can also be found in Section III. Concluding remarks can be found in the last 
Section. 
    
\section{Negatively charged five-body muonic ions}

As mentioned above the energy spectrum of each of the five-body negatively charged ions $a b \mu e_2$, where $(a, b) = (p, d, t)$, contains only one bound electron state, which 
is the ground $1^1S$ state. The five-body negatively charged ion $a b \mu e_2$ includes the central heavy nucleus $a b \mu$ which is positively charged and two bound electrons.
This follows from the fact that the radius of the central nucleus is in $\approx \Big(\frac{m_{\mu}}{m_e}\Bigl) \approx 206.768$ times smaller than the radius of the 
$1s-$electron orbit in the hydrogen atom. It should be mentioned that each of the three-body $(a b \mu)^{+}$ ions has a few different bound states, e.g., each of the $pp\mu, 
pd\mu$ and $pt\mu$ ions has two bound states, each of the $dd\mu$ and $dt\mu$ ions has five bound states, etc (for more detail, see \cite{Fro2011}). Therefore, in experiments 
we can expect the appearance of a few different series of bound states in the five- and six-body muonic systems $a b \mu e_2$ and $a b c \mu e_2$, respectively. In general, 
each bound state in the three-particle quasi-nucleus $(a b \mu)^{+}$ generates one separate series of bound states in the $a b \mu e_2$ and $a b c \mu e_2$ systems. Below, such 
series of bound states are called the fundamental series of bound states. As follows from here the bound state spectra of the $a b \mu e_2$ (= $(a b \mu e_2)^{-}$ ions is 
represented as a combiation of a few fundamental series. If the particles $a$ and $b$ are the bare nuclei of the hydrogen isotopes, then each fundamental series of bound states 
in the $a b \mu e_2$ ion contains only one bound $1^1S-$state.   

To determine the accurate solutions of the non-relativistic Schr\"{o}dinger equation $H \Psi = E \Psi$ (where $E < 0$) in this study we apply the variational expansion of the 
wave function $\Psi$ written in multi-dimensional gaussoids in the relative coordinates $r_{ij} = \mid {\bf r}_j - {\bf r}_k \mid = r_{kj}$ \cite{KT}, \cite{Fro2008}. The symbol 
${\bf r}_i$ designates the Cartesian coordinates of the $i-$th particle. Each of these gaussoids explicitly depends upon a complete set of the relative coordinates $r_{ij}$. For 
the five-particle $a b \mu e_2$ ions one finds ten relative coordinates $r_{ij} = r_{12}, r_{13}, r_{14}, r_{15}, r_{23}, \ldots, r_{45}$. Formally, only nine of these relative 
coordinates are truly independent, but this fact does not complicate analytical operations and numerical computations with the use of multi-dimensional gaussoids (in contrast 
with the basis of linear exponents). 

The variational expansion of the ground state wave function of the $a b \mu e_2$ ion, where $a \ne b$, written in multi-dimensional gaussoids takes the form: 
\begin{eqnarray}
  \Psi = \sum^{N_A}_{i=1} C_i (1 + \hat{P}_{12}) [\exp(-\sum_{(jk)} \alpha^{(i)}_{jk} r^2_{jk})] \; \; \; \label{eq3}
\end{eqnarray}
where $N_A$ is the total number of basis functions used in variational computations, $C_i$ are the linear parameters of this variational expansion, while $\alpha^{(i)}_{jk}$ are the 
non-linear parameters which must be varied in actual calculations. The internal sum in the exponent in Eq.(\ref{eq3}) is calculated over all possible pairs of particles, i.e. $(jk) = 
(12), (13), \ldots, (45)$, where we used the fact that the $(jk)$ and $(kj)$ permutations are identical, i.e., $(jk) = (kj)$. Note that the wave functions, Eq.(\ref{eq3}), 
corresponds to the spatial part of the total wave function. Indeed, this wave function does not contain any spin functions, but it has the correct permutation symmetry in respect to the 
two electron coordinates (particles 1 and 2 in our notation).

In the $p p \mu e_2, d d \mu e_2$ and $t t \mu e_2$ ions we have an additional pair of indistinguishable particles. Antisymmetrization of the total wave functions for such ions is 
slightly complicated. For the spatial part of the total wave function of the $a a \mu e_2$ ion we can write in our notations:
\begin{eqnarray}
  \Psi = \sum^{N_A}_{i=1} C_i (1 + \hat{P}_{12}) (1 + \hat{P}_{45}) [\exp(-\sum_{(jk)} \alpha^{(i)}_{jk} r^2_{jk})] \; \; \; \label{eq35}
\end{eqnarray}
where $\Psi$ is the spatial part of the total wave function of this ion. Since the spin part of the total wave function is a spatial constant, then the spatial wave function $\Psi$ can 
be considered as the total wave function of the $a a \mu e_2$ ion. The basis wave function $\psi_{i}$ in Eq.(\ref{eq3}) is written in the form 
\begin{eqnarray}
 \psi_{i} = \exp(-\alpha^{(i)}_{12} r^2_{12} &-&\alpha^{(i)}_{13} r^2_{13} -\alpha^{(i)}_{14} r^2_{14} -\alpha^{(i)}_{15} r^2_{15} -\alpha^{(i)}_{23} r^2_{23}
 -\alpha^{(i)}_{24} r^2_{24} -\alpha^{(i)}_{25} r^2_{25} \; \; \; \label{eq4} \\ 
 &-&\alpha^{(i)}_{34} r^2_{34} -\alpha^{(i)}_{35} r^2_{35} -\alpha^{(i)}_{45} r^2_{45}) \nonumber
\end{eqnarray}
Correspondingly, the $\hat{P}_{12} \psi_{i}$ function takes the from:
\begin{eqnarray}
 \hat{P}_{12} \psi_{i} = \exp(-\alpha^{(i)}_{12} r^2_{12} &-&\alpha^{(i)}_{23} r^2_{13} -\alpha^{(i)}_{24} r^2_{14} -\alpha^{(i)}_{25} r^2_{15} -\alpha^{(i)}_{13} r^2_{23}
 -\alpha^{(i)}_{14} r^2_{24} -\alpha^{(i)}_{15} r^2_{25}  \; \; \; \label{eq45} \\
 &-&\alpha^{(i)}_{34} r^2_{34} -\alpha^{(i)}_{35} r^2_{35} -\alpha^{(i)}_{45} r^2_{45}) \nonumber
\end{eqnarray}
As follows from these two formulas the $\hat{P}_{12} \psi_{i}$ function can be obtained from the $\psi_{i}$ function by replacing the corresponding non-linear parameters in the exponent 
of the $\psi_{i}$ basis function. The $\hat{P}_{45} \psi_{i}$ and $\hat{P}_{12} \hat{P}_{45} \psi_{i}$ functions are determined analogously:
\begin{eqnarray}
 \hat{P}_{45} \psi_{i} = \exp(-\alpha^{(i)}_{12} r^2_{12} &-&\alpha^{(i)}_{13} r^2_{13} -\alpha^{(i)}_{15} r^2_{14} -\alpha^{(i)}_{14} r^2_{15} -\alpha^{(i)}_{23} r^2_{23}
 -\alpha^{(i)}_{25} r^2_{24} -\alpha^{(i)}_{24} r^2_{25} \; \; \; \label{eq47} \\ 
 &-&\alpha^{(i)}_{35} r^2_{34} -\alpha^{(i)}_{34} r^2_{35} -\alpha^{(i)}_{45} r^2_{45}) \nonumber
\end{eqnarray}
and 
\begin{eqnarray}
 \hat{P}_{12} \hat{P}_{45} \psi_{i} = \exp(-\alpha^{(i)}_{12} r^2_{12} &-&\alpha^{(i)}_{23} r^2_{13} -\alpha^{(i)}_{25} r^2_{14} -\alpha^{(i)}_{24} r^2_{15} -\alpha^{(i)}_{13} r^2_{23}
 -\alpha^{(i)}_{15} r^2_{24} -\alpha^{(i)}_{14} r^2_{25} \; \; \; \label{eq49} \\ 
 &-&\alpha^{(i)}_{35} r^2_{34} -\alpha^{(i)}_{34} r^2_{35} -\alpha^{(i)}_{45} r^2_{45}) \nonumber
\end{eqnarray}

By performing a careful optimization of all ten non-linear parameters in each basis function $\psi_{i}$, where $i$ = 1, $\ldots, N$, one finds an accurate approximation (for relatively 
large $N$) to the spatial part of the actual wave function of an arbitrary $a b \mu e_2$ ion, where $(a, b) = (p, d, t)$. Results of our calculations of the five-body negatively charged 
ions $a b \mu e_2 = (a b \mu e_2)^{-}$ can be found in Table I. Table I contains the total energies of the $a b \mu e_2$ ions. As follows from a comparison of the total energies from 
Tables I and II all five-particle ions considered in this study are stable. A number of bound state properties of these ions (expectation values) can be found in Table III. All these 
properties and energies are expressed in atomic units. Note that almost all electronic properties of the five-body $a b \mu e_2$ ions coincide very well with the expectation values known 
for the negatively charged hydrogen ion(s) H$^{-}$ \cite{Fro2015} (see Table IV). This indicates clearly that the electronic structure of the five-body $a b \mu e_2$ ions is similar to 
the electronic structure of the two-electron H$^{-}$ ion(s). 

\section{Six-body muonic quasi-molecules}

The neutral six-body systems $p d t \mu e_2, d d t \mu e_2, d d d \mu e_2$ and $d t t \mu e_2$, etc, include two heavy positively charged quasi-nucleus, e.g., $dt\mu$ (or $(dt\mu)^{+}$) 
and $d$ (or $d^{+}$), and two bound electrons $e^{-}$. Therefore, the internal structure of these six-body muonic systems must be similar to the two-electron hydrogen H$_{2}$ molecule. 
To respect this fact below we shall call and consider these six-body systems as `quasi-molecules'. As mentioned above the radius of the three-particle `muonic' quasi-nucleus $(dt\mu)^{+}$ 
(and analogous $(pd\mu)^{+}, (pt\mu)^{+}, (dd\mu)^{+}$, etc quasi-nuclei) is in $\approx \Bigl(\frac{m_{\mu}}{m_{e}}\Bigr) \approx 206.768$ times smaller than atomic radius of the hydrogen 
atom which equals $a_0 = \frac{\hbar^2}{m_e e^2} \approx 5.292 \cdot 10^{-9}$ $cm$ (Bohr radius). In this Section we use the same system of notations as above, i.e. the indexes 1 and 2 
mean the electrons $e^{-}$, the index 3 stands for the negatively charged muon $\mu^{-}$, while three other indexes (4, 5 and 6) designate three nuclei of hydrogen isotopes. Note that our 
notation $a b c \mu e_2$ used here contains some uncertainty, since the heaviest quasi-nucleus (or quasi-nucleus which includes the $\mu^{-}-$muon) can be either $a b \mu$, or $a c \mu$, 
or $b c \mu$. Below, we assume that the $b c \mu$ (or $(b c \mu)^{+}$) quasi-nucleus always include the two heaviest nuclei of hydrogen isotopes. This means that the $p d t \mu e_2$ 
system contains the two quasi-nuclei: three-particle $(d t \mu)^{+}$ ion and one bare proton $p^{+}$ (it cannot be the $(p t \mu)^{+}$ ion and bare $d^{+}$ nuclei, or the $(p d \mu)^{+}$ 
ion and bare $t^{+}$ nuclei). The two-electron systems, which can be designated as $d p t \mu e_2$ and $t p d \mu e_2$, with the two other pairs of possible quasi-nuclei: $(p t \mu)^{+}$ 
+ $d^{+}$ and $(p d \mu)^{+}$ + $t^{+}$ are also stable (or quasi-stable), but the absolute values of their total energies are smaller than the total energy of the $p(dt)\mu e_2$ system 
and they are not considered in this study. In other words, our notation $a b c \mu e_2$ always means that the inequality $M_a \le M_b \le M_c$ is obeyed for three nuclear masses of the 
hydrogen isotopes $a, b, c$. 

For the non-symmetric six-body quasi-molecules, e.g., for $p d t \mu e_2$, the trial variational wave function of the correct permutation symmetry takes the form
\begin{eqnarray}
  \Psi = \sum^{N_A}_{i=1} C_i (1 + \hat{P}_{12}) [\exp(-\sum_{(jk)} \alpha^{(i)}_{jk} r^2_{jk})] \; \; \; \label{equ35}
\end{eqnarray}
while for the `partially' symmetric quasi-molecules, e.g., for $d d t \mu e_2$, we can write 
\begin{eqnarray}
  \Psi = \sum^{N_A}_{i=1} C_i (1 + \hat{P}_{12}) (1 + \hat{P}_{45}) [\exp(-\sum_{(jk)} \alpha^{(i)}_{jk} r^2_{jk})] \; \; \; \label{equ351}
\end{eqnarray}
where the internal sum is taken over all 15 permutations of particles $(jk) = (kj) = (12), (13), (14), \ldots$, (45), (46) and (56). Note again that for an arbitrary six-body system we 
have fifteen relative coordinates $r_{ij}$, but in our three-dimensional space only twelve of them are truly independent. Nevertheless, as it was shown in \cite{KT} one can operate with 
the variational expansion, Eq.(\ref{equ351}), by assuming that all fifteen relative coordinates are independent of each other and all variational parameters $\alpha^{(i)}_{jk}$ can be 
varied without any additional restriction. 

As follows from Eqs.(\ref{equ35}) and (\ref{equ351}) the explicit construction of the trial wave functions for both these cases is relatively simple. A real complication can be found for 
the $p p p \mu e_2, d d d \mu e_2$ and $t t t \mu e_2$ quasi-molecules which contain three identical hydrogen nuclei. In these cases we cannot represent the actual wave function as a 
single product of the spatial and spin functions. The actual wave functions of these six-body quasi-molecules must be represented as finite sums of products of the different spatial and 
spin functions. To illustrate this fact let us consider the six-body $p p p \mu e_2$ quasi-molecule. To produce the trial wave function of the correct permutation symmetry we can act as 
for the three-electron Li-atom \cite{Lars}, \cite{Fro2008}. This means that we write the explicit expression for the wave function in the form
\begin{eqnarray}
  \Psi = \sum^{N_A}_{i=1} C_i (1 + \hat{P}_{12}) {\cal A}_{456} \Bigl\{ [\exp(-\sum_{(jk)} \alpha^{(i)}_{jk} r^2_{jk})] \phi_S(456) \Bigr\} \; \; \; \label{equ352}
\end{eqnarray}
where ${\cal A}_{456}$ is the complete antysymmetrizer for three particles with indexes (or numbers) 4, 5 and 6, i.e. ${\cal A}_{456} = 1 - \hat{P}_{45} - \hat{P}_{46} - \hat{P}_{56} + 
\hat{P}_{456} + \hat{P}_{465}$, where $\hat{P}_{jk}$ and $\hat{P}_{ijk}$ are the permutations of the two and three identical particles, respectively. The notation $\phi_S(456)$ stands for 
the spin function(s) of the three protons. Below we designate single-proton spin function in the following  way: spin-up function is the $\alpha-$function, while spin-down function is the 
$\beta-$function (see, e.g., \cite{Lars}, \cite{LLQ}). In these notations one finds the two following spin functions for three-proton system: $\phi^{(1)}_S(456) = \alpha \beta \alpha 
- \beta \alpha \alpha$ and $\phi^{(2)}_S(456) = 2 \alpha \alpha \beta - \beta \alpha \alpha - \alpha \beta \alpha$, where spin-up (or spin-down) function located at the first place (in 
each product of three spin functions) correspond to the first proton, or particle with the index 4, at the second place represents the second proton with index 5 and at the thrid place 
means the proton with the index 6. Such a system of numbering particles by their location in the formula is called the `indexacion by location'. This means that the notation $\alpha \beta 
\alpha$ designates the $\alpha_4 \beta_5 \alpha_6$ spin funcion, while the analogous $\alpha \alpha \beta$ function denotes the $\alpha_4 \alpha_5 \beta_6$ spin function, etc. To determine 
all matrix elements of the Hamiltonian and overlap matrixes in their final forms we need to perform integration over spin variables of all particles, including two electrons and three 
protons. Details of such an integration and explicit forms of the arising spatial projectors can be found in \cite{Fro2011}. Note that in calculations of the total energy and for large 
number of bound state properties of the $p p p \mu e_2, d d d \mu e_2$ and $t t t \mu e_2$ quasi-molecules we can use only one spin functions, e.g., $\phi^{(1)}_S(456)$. The spatial 
projector in this case is ${\cal P} = \frac{1}{12} (2 + 2 \hat{P}_{45} - \hat{P}_{46} - \hat{P}_{56} - \hat{P}_{456} - \hat{P}_{546})$ \cite{Fro2011}. In general, the use of only one spin 
function substantially simplifies all numerical computations. 

Results of our calculations of the six-body $p d t \mu e_2$ and $d d t \mu e_2$ quasi-molecules in their ground $S(L = 0)-$states can be found in Tables I (total energies) and III 
(properties). Again,we have to note that, e.g., the three-particle $d t \mu$ ion (or $(d t \mu)^{+}$ ion) have five bound states (for more detail, see \cite{Fro2008}) and each of these 
bound states generates a separate series of electron bound states in the $d d t \mu e_2$ quasi-molecules (the `fundamental series', see above). In addition to this there is a number of 
electron excited states and many `vibrational' and `rotational' states in each of these six-body quasi-molecules. Moreover, for the $a b c \mu e_2$ quasi-molecule, where $a \ne b \ne c$, 
one finds three different series of bound state spectra, since one of the central heavy nucleus can be either $(a b \mu)^{+}$, or $(a c \mu)^{+}$, or $(b c \mu)^{+}$. These series of 
bound states form the exchange series of the bound state spectra. The difference between the fundamental and exchange series of bound states is obvious, since the fundamental series of 
bound stateis related to the internal structure of the heaviest quasi-nucleus $(b c \mu)^{+}$, while the exchange series are related with the different combinations of hydrogen isotopes 
in the two heavy quasi-nuclei. It is clear that there is no exchange series of the bound state spectra in the five-body $ab \mu e_2$ ions. For the six-body $p d t \mu e_2$ quasi-molecule 
one finds three different exchange series of bound states, while for the $d d t \mu e_2$ system there are two `exchange' series of bound states. The total number of different seires of 
bound electronic states (fundamental + exchange) equals nine for the $p d t \mu e_2$ quasi-molecule and ten for the $d d t \mu e_2$ quasi-molecule.  

As mentioned above in this study for each $a b c \mu e_2$ system we investigate the ground $S(L = 0)-$state only. This bound state has the lowest energy among all possible bound states. 
The computed total energies for all ten $a b c \mu e_2$ systems are presented in Table I in atomic units. Table III contains some bound state properties of the six-body $p d t \mu e_2$ 
quasi-molecule which must be compared with the bound state (electronic) properties of the two-electron hydrogen ${}^{1}$H$_{2}$ (or $ppee$) molecule (see Table IV). As follows from
such a comparison the basic electronic properties of the six-body $p d t \mu e_2$ quasi-molecule (and, in general, any six-body $a b c \mu e_2$ system) are very similar to the analogous 
properties of the two-center H$_{2}$ molecules. Therefore, the electronic structures of these two classes of few-body systems are also similar. This fact must simplify future analysis 
of the six-body $a b c \mu e_2$ systems which contain three nuclei of the hydrogen isotopes. 

\section{Conclusion}

We have considered the bound state spectra of the five- and six-body two-electron muonic systems $a b \mu e_2$ [or $(a b \mu e_2)^{-}$] and $a b c \mu e_2$ which includes the two bound 
electrons, one negatively charged muon $\mu^{-}$ and two (or three) nuclei of hydrogen isotopes, i.e., protium $p$, deuterium $d$ and tritium $t$. Analysis of the bound state properties 
leads us to the conclusion about similarity between the internal structures of the five-body, two-electron ions $a b \mu e_2$ [or $(a b \mu e_2)^{-}]$ and the two-electron H$^{-}$ ion
(three-body system). In particular, this fact follows from an accurate coincidence of a number of basic geometrical and physical properties of the five-body, two-electron ions 
$a b \mu e_2$ with the analogous propeties of the three-body hydrogeen ion H$^{-}$. Such a similarity allows one to predict that each of the $(a b \mu e_2)^{-}$ ions has only one bound 
electron state per one bound state of the central $(a b \mu)^{+}$ ion. 

The bound state spectra in the six-body $a b c \mu e_2$ quasi-molecules are also considered. It is shown that the internal structure of each of these six-body quasi-molecules is similar to 
the electronic structure of the two-center hydrogen H$_2$ molecule. In contrast with the five-body $(a b \mu e_2)^{-}$ ions, each of the six-body quasi-molecules $a b c \mu e_2$ has many 
dozens of electron bound states. Classification of these bound states is a complex problem, which can be solved by considering separation of the bound state spectra into different series of 
bound states, e.g., into the fundamental and exchange series. Accurate and complete analysis of the bound state spectra in the six-body $a b c \mu e_2$ quasi-molecules will reqiure a 
number of years.  In addition to this, now we can determined a large number of expectation values (i.e. bound state properties) of these quasi-molecules to a good accuracy. 

In conclusion, we note again that the direct numerical calculations of the bound state spectra in the five- and six-body two-electron muonic systems $a b \mu e_2$ [or $(a b \mu e_2)^{-}$] 
and $a b c \mu e_2$ is a new and significant step in our understanding of the internal sructure of these systems and processes in them. In one short paper it is impossible to answer all
important questions. The next goal in studies of five- and six-body two-electron muonic systems $a b \mu e_2$ [or $(a b \mu e_2)^{-}$] and $a b c \mu e_2$ is to increase the overall 
accuracy of numerical computations of bound states in these ions and/or quasi-molecules. In addition to this, we need to determine more bound state properties, including the expectation 
values of all interparticle delta-functions, triple delta-functions, etc. These values are needed to predict the actual rates of different processes, e.g., nuclear fusion rate(s), in 
different five- and six-body two-electron muonic systems. In particular, the study of the  six-body $a b c \mu e_2$ quasi-molecules which contain either the $dt\mu$ quasi-nucleus, or 
$dd\mu$ quasi-nucleus in their weakly bound states (the excited $L$-states, or (1,1)-states) is of great theoretical interest, since in these quasi-nucleus we can observe a strong mixture 
of the `nuclear' and electron bound states. In other words, the electron and nuclear motions in such six-body systems cannnot be separated from each other and must be considered together. 
Even classification of such `mixed' bound states is a very interesting problem. 
  
\section{Acknowledgments}

This work was supported in part by the NSF through a grant for the Institute for Theoretical Atomic, Molecular, and Optical Physics (ITAMP) at Harvard University and the Smithsonian 
Astrophysical Observatory. Also, I wish to thank James Babb (ITAMP, Cambridge, MA) and Leonid I. Menshikov (Kurchatov Institute, Moscow, Russia) for stimulating discussions.

\newpage
\begin{table}[tbp]
   \caption{Total energies in atomic units (in $a.u.$) of the ground $S(L = 0)-$states in the five-body hydrogen-muonic ions $(a b \mu e_2)^{-}$ and 
             six-body quasi-molecules $a b c \mu e_2$.}
     \begin{center}
     \begin{tabular}{| c | c | c | c |}
      \hline\hline
 $N$ &  $p d \mu e_2$ & $p t \mu e_2$ & $d t \mu e_2$ \\ 
     \hline
 200 & -106.53518185 & -108.01709125 & -111.88679470 \\
         \hline\hline
 $N$  & $p p \mu e_2$ & $d d \mu e_2$ & $t t \mu e_2$ \\ 
     \hline
 200 & -102.7500577 & -110.34333639 & -113.49897897 \\
       \hline\hline
 $N$  & $p d t \mu e_2$ & $p p d \mu e_2$ & $p p t \mu e_2$ \\ 
     \hline
 200 & -112.47480351 & -107.12363558 & -108.60558047 \\
         \hline 
 $N$ & $d d t \mu e_2$ & $p d d \mu e_2$ & $p t t \mu e_2$ \\ 
     \hline
 200 & -112.47116785 & -110.94095171 & -114.10055093$^{(a)}$ \\
         \hline 
 $N$  & $p p p \mu e_2$ & $d d d \mu e_2$ & $t t t \mu e_2$ \\ 
     \hline
 200 & -103.33587577 & -110.92797031 & -114.07956931 \\
         \hline \hline  
  \end{tabular}
  \end{center}
 ${}^{(a)}$The total energy of the ground bound state of the $d t t \mu e_2$ system is -114.10115887$a.u.$
  \end{table}
\begin{table}[tbp]
   \caption{Total energies in atomic units (in $a.u.$) of the ground $S(L = 0)-$states of the three-body muonic molecular ions $a b \mu$ (or $(a b \mu)^{+}$).
            Numerical values of all total energies have been taken from \cite{Fro2011}, where all particle masses are exactly the same as they used 
            in this study.} 
     \begin{center}
     \begin{tabular}{| c | c | c | c |}
      \hline\hline
 $N$ &  $p p \mu$ & $d d \mu$ & $t t \mu$ \\ 
     \hline
 $E$ & -102.2235035785787 & -109.8169263959981 & -112.9728490317648 \\
         \hline\hline
 $N$ &  $p d \mu$ & $p t \mu$ & $d t \mu$ \\ 
     \hline
 $E$ & -106.0125270695158 & -107.4947026128185 & -111.3643469153818 \\ 
         \hline \hline  
  \end{tabular}
  \end{center}
  \end{table}
%
%
 \begin{table}[tbp]
   \caption{The expectation values of a number of properties (in $a.u.$) of the ground $S(L = 0)-$states in some five-body $(a b \mu e_2)^{-}$ ions 
            and six-body $a b c \mu e_2$ quasi-molecules.}
     \begin{center}
     \begin{tabular}{| c | c | c | c | c |}
       \hline\hline          
 ion/molecule  & $\langle r^{-2}_{p \mu} \rangle$ & $\langle r^{-1}_{p \mu} \rangle$ & $\langle r_{p \mu} \rangle$ & $\langle r^2_{p \mu} \rangle$ \\
     \hline
   $p d \mu e_2$ & 52465.736 & 155.77373 & 0.01009561 & 0.000137855 \\
       \hline\hline  
 ion/molecule  & $\langle r^{-2}_{d \mu} \rangle$ & $\langle r^{-1}_{d \mu} \rangle$ & $\langle r_{d \mu} \rangle$ & $\langle r^2_{d \mu} \rangle$ \\
       \hline\hline    
   $d d \mu e_2$ & 48961.114 & 150.62685 & 0.01025263 & 0.001390777 \\

 $p d t \mu e_2$ & 47948.92  & 149.4191  & 0.0102416  & 0.00013845  \\ 
     \hline\hline     
 ion/molecule  & $\langle r^{-2}_{d e} \rangle$ & $\langle r^{-1}_{d e} \rangle$ & $\langle r_{d e} \rangle$ & $\langle r^2_{d e} \rangle$ \\
     \hline
   $p d \mu e_2$ & 1.1008758 & 0.68634114 & 2.5693953 & 9.8722965 \\
  
   $d d \mu e_2$ & 1.1104775 & 0.68433819 & 2.6485125 & 10.895838 \\ 

 $p d t \mu e_2$ & 1.488335  & 0.8834730  & 1.606050  & 3.248165  \\
     \hline\hline     
 ion/molecule  & $\langle r^{-2}_{ee} \rangle$ & $\langle r^{-1}_{ee} \rangle$ & $\langle r_{ee} \rangle$ & $\langle r^2_{ee} \rangle$ \\
     \hline
   $p d \mu e_2$ & 0.1612780 & 0.3201021 & 4.1271305 & 20.8879 \\
  
   $d d \mu e_2$ & 0.1573468 & 0.3142697 & 4.2891434 & 23.0935 \\

 $p d t \mu e_2$ & 0.507390  & 0.574317  & 2.220188  & 5.85409 \\
     \hline\hline     
 ion/molecule  & $\langle \frac12 {\bf p}^2_{e} \rangle$ & $\langle \frac12 {\bf p}^2_{\mu} \rangle$ & $\langle \frac12 {\bf p}^2_{p} \rangle$ & $\langle \frac12 {\bf p}^2_{d} \rangle$ \\ 
     \hline
   $p d \mu e_2$ & 0.263352470 & 19683.345 & 15710.399 & 11997.327 \\
  
   $d d \mu e_2$ & 0.263570803 & 20880.742 & --------  & 16204.694 \\ 

 $p d t \mu e_2$ & 0.22206238  & 21510.84  & 23875.25  & 21291.87  \\
      \hline\hline
  \end{tabular}
  \end{center}
 ${}^{(a)}$The $\langle \frac12 {\bf p}^2_{t} \rangle$ expectation value for the $p d t \mu e_2$ quasi-molecule is $\approx$ 20357.74.
   \end{table}
%

 \begin{table}[tbp]
   \caption{The expectation values of a number of properties (in $a.u.$) of the ground states in the H$^{-}$ ion (or ${}^{\infty}$H$^{-}$ ion) and ${}^{1}$H$_{2}$ (or $ppee$) molecule.
            The notation $e$ stands for the electron(s), while $N$ designates the heavy nucleus (or nuclei).}
     \begin{center}
     \scalebox{0.85}{%
     \begin{tabular}{| c | c | c | c | c |}
       \hline\hline          
 ion/molecule  & $\langle r^{-2}_{N e} \rangle$ & $\langle r^{-1}_{N e} \rangle$ & $\langle r_{N e} \rangle$ & $\langle r^2_{N e} \rangle$ \\
     \hline
           H$^{-}$ & 1.11666282452542572 & 0.6832617676515272224 & 2.7101782784444203653 & 11.913699678051262274 \\
  
   ${}^{1}$H$_{2}$ & 1.5730801 & 0.9017345 & 1.5751544 & 3.146458 \\
     \hline    
  ion/molecule  & $\langle r^{-2}_{e e} \rangle$ & $\langle r^{-1}_{e e} \rangle$ & $\langle r_{e e} \rangle$ & $\langle r^2_{e e} \rangle$ \\
     \hline
           H$^{-}$ & 0.15510415256242466 & 0.311021502214300052 & 4.4126944979917277211 & 25.202025291240331897 \\
  
   ${}^{1}$H$_{2}$ & 0.5050168 & 0.5793405 & 2.2012319 & 0.5803927 \\
         \hline 
  ion/molecule  & $\langle r^{-2}_{N N} \rangle$ & $\langle r^{-1}_{N N} \rangle$ & $\langle r_{N N} \rangle$ & $\langle r^2_{N N} \rangle$ \\
          \hline 
   ${}^{1}$H$_{2}$ & 0.497428 & 0.699864 & 1.450543 & 2.135437 \\  
      \hline\hline
  \end{tabular}}
  \end{center}
 ${}^{(a)}$The $\langle \frac12 {\bf p}^2_{e} \rangle$ expectation value for the ${}^{\infty}$H$^{-}$ ion (computed with the same wave function) 
           is $\approx$ 0.2638755082721885983 $a.u.$ 
   \end{table}

\begin{thebibliography}{01}

\bibitem{Sachar} Ya.B. Zeldovich and A.D. Sakharov, JETP {\bf 32}, 947 (1957) [Sov. Phys. JETP {\bf 5}, 775 (1957)].

\bibitem{Alv} L.W. Alvarez, H. Bradner, F.S. Crawford, Jr., J.A. Crawford, P. Falk-Vairant, M.L. Good, J.D. Gow, A.H. Rosenfeld, 
F. Solmitz, M.L. Stevenson, H.K. Ticho, and R.D. Tripp, Phys. Rev. {\bf 105}, 1127 (1957). 

\bibitem{Zeld} Ya.B. Zeldovich and S.S. Gershtein, Usp. Fiz. Nauk {\bf 71} 581 (1960) [Sov. Phys. Usp. {\bf 3}, 593 (1961)]. 

\bibitem{Jones} S.E. Jones, Nature {\bf 321} 127 (1986).  

\bibitem{Advc} K. Langanke and C.A. Barnes, {\it Nucleosynthesis in the Big Bang and in Stars}, in {\it Advances in Nuclear Physics} 
(eds. J.W. Negie and E. Wogt, Spinger Verlag, Berlin, 1998), Chpt. 5.  

\bibitem{MS} L.I. Menshikov and L.N. Somov, Sov. Phys. Usp. {\bf 33}, 616 (1990).

\bibitem{Fro2011} A.M. Frolov and David M. Wardlaw, Europ. Phys. Journal D {\bf 63}, 339 (2011); ibid, {\bf 66}, 212 (2012).  

\bibitem{Fro2015} A.M. Frolov, Europ. Phys. Journal D {\bf 69}, 132 (2015).

\bibitem{KT} N.N. Kolesnikov and V.I. Tarasov, Yad. Fiz. \textbf{35}, 609 (1982), [Sov. J. Nucl. Phys. \textbf{35}, 354 (1982)].

\bibitem{Lars} S. Larsson, Phys. Rev. \textbf{169}, 49 (1968).

\bibitem{Fro2008} A.M. Frolov and D.M. Wardlaw, Phys. Rev. A Phys. Rev. A \textbf{78}, 042506 (2008); JETP {\bf 138}, 5 (2010).

\bibitem{LLQ} L.D. Landau and E.M. Lifshitz, {\it Quantum Mechanics: Non-Relativistic Theory}, (3rd. ed. Pergamon Press, New York (1976)).

\end{thebibliography}
\end{document}